\documentclass[11pt]{amsart}
\numberwithin{equation}{section}
\begin{document}

\title[]{A four-dimensional Hooke's law can encompass linear
    elasticity and inertia} \thanks{Nuovo Cimento B, in press}
\author{S. Antoci \and L. Mihich}
\address{ Dipartimento di Fisica ``A. Volta'' and I. N. F. M.,
    Via Bassi 6, Pavia, Italy}
\email{Antoci@fisav.unipv.it} 

\begin{abstract}
The question is examined, whether the formally straightforward
    extension of Hooke's time-honoured stress-strain relation to the
    four dimensions of special and of general relativity can make
    physical sense. The four-dimensional Hooke's law is found able to account
    for the inertia of matter; in the flat space, slow motion approximation
    the field equations for the ``displacement'' four-vector field $\xi^i$
    can encompass both linear elasticity and inertia. In this limit
    one just recovers the equations of motion of the classical theory
    of elasticity.
\end{abstract}
\maketitle

    \section{Introduction}
    After having fostered the birth of the four-dimensional
    approach to special relativity~\cite{ref:Minkowski},
    macroscopic electromagnetism has found its natural expression
    in the four dimensional language of general relativity,
    both {\it in vacuo}~\cite{ref:Einstein} and in
    matter~\cite{ref:Nordstroem},~\cite{ref:Gordon}. Its field
    quantities and its field equations have achieved a canonical
    form, that can be summarized as follows~\cite{ref:Post}:
    the unconnected space-time manifold suffices for writing Maxwell's
    equation in the naturally invariant form:
    \begin{equation}\label{1.1}
    {\bf H}^{ik}_{~~,k}={\bf s}^{i},
    \end{equation}
    \begin{equation}\label{1.2}
    F_{[ik,m]}=0,
    \end{equation}
    where ${\bf H}^{ik}$ is a skew, contravariant tensor density that
    represents the electric displacement and the magnetic field, while
    $F_{ik}$ is a covariant skew tensor that accounts for the electric
    field and for the magnetic induction. Even if the four-current density
    ${\bf s}^i$ is given and the co-ordinate system is fixed, these
    equations are not sufficient for determining both ${\bf H}^{ik}$
    and $F_{ik}$. They fulfil the two identities ${\bf H}^{ik}_{~~,k,i}=0$,
    which ensures the conservation of the electric four-current, and
    ${\bf e}^{ikmn}F_{[ik,m],n}=0$, where ${\bf e}^{ikmn}$ is the totally
    antisymmetric tensor density of Ricci and Levi Civita. By simply counting
    the components of the fields, the equations and the identities one
    gathers that (\ref{1.1}) and (\ref{1.2}) need to be complemented
    by the so-called constitutive equation, {\it i.~e.} by some tensor
    equation that uniquely defines for instance ${\bf H}^{ik}$ in
    terms of $F_{ik}$ and of whatever fields may be needed in
    describing the features of the electromagnetic medium. For a linear
    medium the constitutive equations can be written as~\cite{ref:Nordstroem}:
    \begin{equation}\label{1.3}
    {\bf H}^{ik}={\frac{1}{2}}{\bf X}^{ikmn}F_{mn};
    \end{equation}
    the properties of the medium are specified by the four-index
    tensor density ${\bf X}^{ikmn}$. To complete this schematic
    picture of macroscopic electromagnetism, one may add that,
    once the constitutive equation is given, Maxwell's equations
    can be solved in terms of the four-vector potential $\varphi_i$.
    If $F_{ik}$ is defined as:
    \begin{equation}\label{1.4}
    F_{ik}\equiv\varphi_{k,i}-\varphi_{i,k},
    \end{equation}
    equation (\ref{1.2}) is satisfied. Then the constitutive equation
    (\ref{1.3}) takes the form:
    \begin{equation}\label{1.5}
    {\bf H}^{ik}={\frac{1}{2}}{\bf X}^{ikmn}
    (\varphi_{n,m}-\varphi_{m,n})
    \end{equation}
    and the four components of $\varphi_i$ are determined, up to a
    gauge transformation, by the four equations (\ref{1.1}).
    One may well wonder why the introduction to a paper announced
    to deal with elasticity and inertia begins with these
    electromagnetic reminiscences. But equation (\ref{1.5}),
    this pervasive, general relativistic, four-dimensional
    constitutive equation is the antisymmetric counterpart
    of an equally pervasive, although pre-relativistic,
    three-dimensional and symmetric constitutive equation:
    \begin{equation}\label{1.6}
    {\Theta}^{\lambda\mu}={\frac{1}{2}}C^{\lambda\mu\rho\sigma}
    (\xi_{\rho,\sigma}+\xi_{\sigma,\rho}).
    \end{equation}
    This is the law of linear elasticity, disclosed~\cite{ref:Hooke}
    by Hooke in 1678 with the words ``{\it ut~tensio sic vis}''.
    It is given here by using the language of three-dimensional
    tensors and of tensor analysis, that were just invented to
    cope with the far-reaching developments stemmed from Hooke's
    discovery. In this way all the analogies between (\ref{1.5})
    and (\ref{1.6}) are made apparent: the four-potential
    $\varphi_i(x^k)$ is the four-dimensional counterpart of the
    displacement vector $\xi_{\rho}(x^\lambda)$, the deformation
    tensor
    \begin{equation}\label{1.7}
    S_{\rho\sigma}\equiv{\frac{1}{2}}
    (\xi_{\rho,\sigma}+\xi_{\sigma,\rho})
    \end{equation}
    is the symmetric, three-dimensional counterpart of $F_{ik}$,
    while the symmetric, three-dimensional stress tensor
    ${\Theta}^{\lambda\mu}$ of (\ref{1.6}) is replaced in (\ref{1.5})
    with the skew, four-dimensional tensor density ${\bf H}^{ik}$.
    One might indulge in disclosing further analogies
    that betray the mechanistic origin of electromagnetism, but a
    question comes to the mind: have equation (\ref{1.6}) and the
    classical theory of elasticity found a generally accepted
    reformulation in the framework of special and of general
    relativity, as it has occurred with electromagnetism? Even a
    cursory inspection of the literature shows that the answer is
    negative. This problem has attracted considerable attention
    in the years soon after 1905, as it is testified {\it e.~g.} by
    the seminal paper of Herglotz on the mechanics of deformable
    bodies~\cite{ref:Herglotz}. After 1915 Nordstr\"om
    resumed~\cite{ref:Nordstroem} Herglotz' approach and translated
    it in a general relativistic form, but, to our knowledge, the
    interest in the issue sinked, and only resurfaced at the time
    when Weber undertook~\cite{ref:Weber} his studies on the detection of
    gravitational waves. In classical elasticity, the deformation
    of an elastic body is measured relative to a natural unstrained
    state. Synge found it difficult to carry over this idea into a
    pseudo-Riemannian space-time and decided~\cite{ref:Synge} that
    the sentence ``rate of change of stress linear function of rate of
    strain'' should replace, in general relativity, Hooke's Latin
    dictum. Rayner resurrected the concept of reference state under
    the form of a reference metric~\cite{ref:Rayner} and postulated the
    following relativistic variant of Hooke's law:
    \begin{equation}\label{1.8}
    \Sigma_{ij}\equiv{\frac{1}{2}}
    c_{ij}^{~~~kl}(\gamma_{kl}-\gamma^0_{kl})
    \end{equation}
    where the auxiliary metric
    \begin{equation}\label{1.9}
    \gamma^{ik}=g^{ik}+u^iu^k
    \end{equation}
    defined in terms of the true metric $g_{ik}$ and of the
    four-velocity $u^i$, and its reference counterpart
    $\gamma^0_{ik}$ are availed of. Despite its four-dimensional
    clothing, equation (\ref{1.8}) defines an effectively
    three-dimensional stress tensor: since both the auxiliary
    metric and its reference countepart are orthogonal to the
    four-velocity, one finds
    \begin{equation}\label{1.10}
    \Sigma_{ij}u^j\equiv 0,
    \end{equation}
    as it is appropriate for describing the elastic stress. Carter
    and Quintana extended Rayner's theory to a non-linear
    regime~\cite{ref:CQ}, as it is necessary when dealing with
    astrophysical situations. Nordstr\"om
    derived~\cite{ref:Nordstroem} his equations of motion from
    Hamilton's principle and showed their coincidence with the ones
    obtained through the general relativistic law of conservation for
    the energy tensor
    \begin{equation}\label{1.11}
    {\bf{T}}^{ik}_{;k}=0.
    \end{equation}
    The later authors mentioned above and many
    others~\cite{ref:Hernandez}-\cite{ref:Lamoureux} instead relied
    only on equation (\ref{1.11}) for setting up their equations of
    motion. There is however a notable exception in the
    Lagrangian formulation given~\cite{ref:KM} by
    Kijowski and Magli.

     \section{A really four-dimensional extension of Hooke's law}
    The diverse proposals for a general relativistic reformulation
    of Hooke's law all adopt an effectively three-dimensional
    stress-strain relation. It seems therefore natural to wonder
    whether a truly four-dimensional Hooke's law can have some
    physical sense. The general relativistic extension of equation
    (\ref{1.6}) is formally immediate: in a pseudo-Riemannian
    space-time whose metric is $g_{ik}$ one considers a contravariant
    four-vector field  $\xi^i(x^k)$ that aims at representing some
    ``displacement'', and uses its covariant counterpart to define
    a four-dimensional ``deformation'' tensor
    \begin{equation}\label{2.1}
    S_{ik}={1\over2}(\xi_{i;k}+\xi_{k;i}).
    \end{equation}
    A four-dimensional ``stiffness'' tensor density ${\bf{C}}^{iklm}$
    is then introduced; it will be symmetric in both the first pair
    and the second pair of indices, since it will be used for
    producing a ``stress-momentum-energy'' tensor density
    \begin{equation}\label{2.2}
    {\bf{T}}^{ik}={\bf{C}}^{iklm}S_{lm},
    \end{equation}
    through the four-dimensional extension of Hooke's law.
    If ${\bf{T}}^{ik}$ is meant to be the overall energy tensor density,
    it must obey the four-dimensional conservation law (\ref{1.11}),
    that of course has to do with the equations of motion of the
    field $\xi^i(x^k)$. From a formal standpoint, this is nearly the
    end of the story. The real question is: can this abstract scheme find
    a physical interpretation, at least in some limit condition?\par
    Let us start inquiring whether the vector field $\xi^i(x^k)$ can
    admit of the following physical meaning. One considers a
    co-ordinate system such that, at a given event:
    \begin{equation}\label{2.3}
    g_{ik}=\eta_{ik}\equiv{diag(1,1,1,-1)},
    \end{equation}
    while the Christoffel symbols are all vanishing,
    and the components of the four-velocity of matter are
    \begin{equation}\label{2.4}
    u^1=u^2=u^3=0,\ u^4=1.
    \end{equation}
    In this co-ordinate system we suppose it possible to measure,
    at the chosen event, both the components of the spatial
    displacement of the elastic medium from some relaxed condition
    and the proper time, that we interpret as temporal
    displacement. If these four quantities can be determined at any event
    with this procedure, they can be used to define, through
    the appropriate transformations, the four-vector
    field $\xi^i(x^k)$ in an arbitrary co-ordinate system. Let us assume
    that the field $\xi^i(x^k)$ that we have previously introduced in a
    purely formal way really allow for this physical interpretation.
    Then, if the material is unstrained in the ordinary sense, the only
    nonvanishing component of $S_{ik}$ should be $S_{44}$, whose
    value should be $-1$. This remark suggests defining the four-velocity
    $u^i$ of matter through the equation
    \begin{equation}\label{2.5}
    \xi^i_{;k}u^k=u^i.
    \end{equation}
    A necessary condition for the above definition to hold is:
    \begin{equation}\label{2.6}
    det(\xi^i_{;k}-\delta^i_k)=0.
    \end{equation}
    This shall be one equation that the field $\xi^i$ must satisfy;
    the number of independent components of $\xi^i$ will thereby
    be reduced to three. We observe that the very definition of a
    four-velocity field $u^i(x^k)$ by starting from the field $\xi^i$
    is not ensured {\it a priori}; its possible existence will
    depend on the physical properties of the model that one considers.\par
    A four-dimensional ``stiffness'' tensor $C^{iklm}$ possibly endowed
    with physical meaning can be built as follows. We assume that in the
    locally Min\-kowskian rest frame defined above the only nonvanishing
    components of $C^{iklm}$ are $C^{\lambda\nu\sigma\tau}$,
    with the tentative r\^{ole} of elastic moduli, and
    \begin{equation}\label{2.7}
    C^{4444}=-\rho,
    \end{equation}
    where $\rho$ measures the rest density of matter (energy). We need
    to write the four-dimensional ``stiffness'' tensor in an arbitrary
    co-ordinate system. The task can be easily accomplished if matter
    is isotropic when looked at in the above mentioned Minkowskian rest frame.
    Just for the sake of simplicity, we shall deal henceforth only with this
    case. One avails of the auxiliary metric (\ref{1.9}); then the part of
    $C^{iklm}$ stemming from the ordinary elasticity of the isotropic
    medium can read~\cite{ref:CL}
    \begin{equation}\label{2.8}
    C^{iklm}_{el.}=-\lambda\gamma^{ik}\gamma^{lm}
    -\mu(\gamma^{il}\gamma^{km}+\gamma^{im}\gamma^{kl}),
    \end{equation}
    where $\lambda$ and $\mu$ are assumed to be constants. The part
    of $C^{iklm}$ accounting for the inertial term reads instead
    \begin{equation}\label{2.9}
    C^{iklm}_{in.}=-\rho u^iu^ku^lu^m.
    \end{equation}
    When no other fields are present, the tensor density ${\bf{T}}^{ik}$
    defined by equation (\ref{2.2}) must fulfil the four
    conditions (\ref{1.11}). Since $u^i$ is defined in terms of $\xi^i$
    and of the metric tensor $g_{ik}$ via equation (\ref{2.5}),
    imposing the four conditions (\ref{1.11}) means postulating four
    nonlinear partial differential equations that, for a given $g_{ik}$,
    must be obeyed by the three independent components of the field
    $\xi^i$ and by the scalar field $\rho$. Thanks to
    (\ref{2.8}) and (\ref{2.5}), the putative elastic part of the
    energy tensor $T^{ik}$ for isotropic matter reads:
    \begin{eqnarray}\label{2.10}
    T^{ik}_{el.}=C^{iklm}_{el.}S_{lm}
    =-\lambda(g^{ik}+u^iu^k)(\xi^m_{;m}-1)\nonumber\\
    -\mu[\xi^{i;k}+\xi^{k;i}+u_l(u^i\xi^{l;k}+u^k\xi^{l;i})],
    \end{eqnarray}
    and is orthogonal to the four-velocity, as an intrinsically
    three-dimensional tensor should be, while the supposedly inertial
    part of $T^{ik}$ turns out to be effectively so. In fact,
    due to equation (\ref{2.5}):
    \begin{equation}\label{2.11}
    T^{ik}_{in.}=C^{iklm}_{in.}S_{lm}
    =\rho u^iu^k.
    \end{equation}
    When $\lambda=\mu=0$ we have no elasticity at all, and the field
    equations (\ref{1.11}) reduce to:
    \begin{equation}\label{2.12}
    \{\rho u^iu^k\}_{;k}
    =\{\rho u^k\}_{;k}u^i
    +\rho u^i_{~;k}u^k=0.
    \end{equation}
    By contracting the last equation with $u_i$ one finds:
    \begin{equation}\label{2.13}
    \{\rho u^k\}_{;k}=0.
    \end{equation}
    This equation is convenient for finding $\rho$, once the
    problem of motion, entirely written in kinematic terms:
    \begin{equation}\label{2.14}
    u^i_{~;k}u^k=0,
    \end{equation}
    has been solved. The world lines of matter are then timelike geodesics
    of the spacetime whose metric is $g_{ik}$. Once the geodesic equations
    are solved for $u^i$ also the displacement field $\xi^i$ is in principle
    determined through the equations (\ref{2.5}) and (\ref{2.6}). In the limit
    case when equation (\ref{2.3}) holds everywhere the world lines of
    matter are straight and by performing a Lorentz transformation we can
    put matter everywhere to rest in the co-ordinate system, say,
    $x^i=(x, y, z, t)$. Then equation (\ref{2.4}) is satisfied everywhere;
    we can fulfil both (\ref{2.5}) and (\ref{2.6}) by posing:
     \begin{equation}\label{2.16}
    \xi^i=(0,0,0,t).
    \end{equation}

    \section{The classical theory of elasticity as limit case}
    Let us hold fixed both the co-ordinate system and the metric selected
    above, and assume that now the constants $\lambda$ and $\mu$
    are nonvanishing, but that $u^i$ differs very slightly from its rest
    form (\ref{2.4}). We deal with the components $u^\rho$ as with
    first-order infinitesimal quantities, while of course the
    increment to $u^4$ shall be infinitesimal at second order.
    Since one expects that $\xi^{\nu}_{,\rho}$, like  $\xi^{\nu}$,
    and $\xi^4_{,\rho}$ will be at most first order infinitesimal
    quantities, the first order approximation to equation
    (\ref{2.5}) requires
    \begin{equation}\label{3.1}
    u^\nu=\xi^\nu_{,4},
    \end{equation}
    and
    \begin{equation}\label{3.2}
    u^4=\xi^4_{,4}=1.
    \end{equation}
    From the above remark about $\xi^i_{,\rho}$ and from equations
    (\ref{3.1}), (\ref{3.2}) one gathers that $\xi^i_{,k}$ will
    fulfil equation (\ref{2.6}) to the required first order. Let
    us check the changes occurred to equation (\ref{2.13}), that
    expressed the conservation of matter when ordinary elasticity
    was absent. One writes:
    \begin{equation}\label{3.3}
    T^{ik}_{~;k}u_i=(T^{ik}u_i)_{;k}-T^{ik}u_{i;k}=0
    \end{equation}
    and, since
    \begin{equation}\label{3.4}
    T^{ik}_{el.}u_k=0,
    \end{equation}
    equation (\ref{3.3}) comes to read:
    \begin{equation}\label{3.5}
    (-\rho u^k)_{;k}-T^{ik}_{el.}u_{i;k}=0.
    \end{equation}
    While the inertial term of this equation is a first-order
    infinitesimal, the putative elastic term contains only infinitesimal
    quantities of a higher order. Therefore equation (\ref{2.13})
    still holds, with the required approximation. We write now the field
    equations (\ref{1.11}) by retaining only the first-order terms.
    The equations for $i=1,2,3$ are:
    \begin{eqnarray}\label{3.6}
    \{\rho u^\nu u^k
    -\lambda(\eta^{\nu k}+u^\nu u^k)(\xi^m_{,m}-1)\nonumber\\
    -\mu[\xi^{\nu,k}+\xi^{k,\nu}
    +u_l(u^\nu\xi^{l,k}+u^k\xi^{l,\nu})]\}_{,k}=0.
    \end{eqnarray}
    Due to (\ref{2.13}) and (\ref{2.5}), one can rewrite these
    equations as
    \begin{equation}\label{3.7}
    \rho u^{\nu}_{~,4}u^4=
    \lambda\xi^{m,\nu}_{~,m}
    +\mu(\xi^{\nu,\rho}+\xi^{\rho,\nu})_{,\rho};
    \end{equation}
    one eventually obtains, to the required order:
    \begin{equation}\label{3.8}
    \rho\xi^\nu_{,4,4}=
    \lambda\xi^{m,\nu}_{~,m}
    +\mu(\xi^{\nu,\rho}+\xi^{\rho,\nu})_{,\rho}.
    \end{equation}
    Equation (\ref{3.2}) says that $\xi^4_{,4}$ can differ from unity
    at most for second order infinitesimal terms; hence, with the
    required accuracy, $\xi^{\rho,\nu}_{,\rho}$ can be substituted for
    $\xi^{m,\nu}_{,m}$ in the previous equations, that come to read:
    \begin{equation}\label{3.9}
    \rho\xi^\nu_{,4,4}=
    \lambda\xi^{\rho,\nu}_{~,\rho}
    +\mu(\xi^{\nu,\rho}+\xi^{\rho,\nu})_{,\rho}.
    \end{equation}
    The field equation for $i=4$ reads:
    \begin{eqnarray}\label{3.10}
    \{\rho u^4 u^k
    -\lambda(\eta^{4k}+u^4 u^k)(\xi^m_{,m}-1)\nonumber\\
    -\mu[\xi^{4,k}+\xi^{k,4}
    +u_l(u^4\xi^{l,k}+u^k\xi^{l,4})]\}_{,k}=0.
    \end{eqnarray}
    The elastic terms provide only second-order contributions;
    therefore only the inertial term survives and, to the required
    order, the last field equation reads
    \begin{equation}\label{3.11}
    \{\rho u^4u^k\}_{,k}=0,
    \end{equation}
    as it occurs for the geodesic motion. The conservation equation
    (\ref{2.13}) could have been adopted as fourth equation as well.
    The three equations (\ref{3.9}) and the conservation
    law~(\ref{2.13}) exactly match the corresponding equations of the
    classical theory of elasticity.

    \section{Conclusion}
    As far as one can understand from the results of the previous
    Section, postulating a truly four-dimensional Hooke's law
    does not seems to be a false step, since it allows to merge
    both linear elasticity and inertia into a sort of extended,
    four-dimensional elasticity. Investigating the content of this
    approach when the metric is highly nonflat and
    the relative speed of different portions of matter is comparable
    with the velocity of light is a mathematically difficult and physically
    useless undertaking, since one must expect that the linear approximation
    of the classical theory of elasticity will fail to account for the
    behaviour of matter under such extreme conditions.
    Writing the equations of motion of elastic matter in
    presence of a weak, wavy deviation of $g_{ik}$ from
    flatness \footnote[1]{The very concept of weak gravitational
    wave is not without pitfalls. As wittily stressed by
    Eddington~\cite{ref:Eddington},
    one can easily manifacture perturbations of the flat metric
    that ``propagate'' with the {\it speed of thought.}} is instead both
    mathematically affordable and slightly more useful: in the
    complete lack of experimental evidence about the way elastic
    matter interacts with gravitational waves, it will at least allow
    for a comparison with the theoretical predictions achievable
    through the diverse approaches recalled in the Introduction.\par
    Confronted with the bareness of the experimental
    landscape, one can however seek consolation in the sense of
    intellectual fulfilment that looking at known
    things from a new perspective always provides. The expectation aroused
    by the analogy between the four-dimensional constitutive equation of
    electromagnetism (\ref{1.5}) and the three-dimensional expression
    (\ref{1.6}) of the stress-strain relation of classical elasticity
    was not void. Viewing inertia as a kind of elasticity is indeed
    possible, and this possibility is not constrained to the weak
    field, slow motion approximation. The reformulation of the inertial
    term of $T^{ik}$ was possible thanks to equations (\ref{2.5}) and
    (\ref{2.6}), whose r\^ole was central in forcing the four-vector field
    $\xi^i$ to account for both inertia and ordinary, classical elasticity.

\bibliographystyle{amsplain}

\end{document}